\documentstyle[12pt,aaspp4]{article}
\def\himsun{{h^{-1}M_\odot}}

\def\pppm{\rm P^3M}
\def\beq{\begin{equation}}
\def\eeq{\end{equation}}
\def\mpc{\,h^{-1}{\rm {Mpc}}}

\begin{document}

\title{The Angular Momentum Distribution within Halos in Different Dark
Matter Models}

\author{D.N.Chen, Y.P. Jing}
\affil{Shanghai Astronomical Observatory, the Partner Group of MPI f\"ur
Astrophysik, Nandan Road 80, Shanghai 200030, China}
\affil{dnchen@center.shao.ac.cn; ypjing@center.shao.ac.cn}
\received{2001 ???}
\accepted{2001 ???}

\begin{abstract}
We study the angular momentum profile of dark matter halos for a
statistical sample drawn from a set of high-resolution cosmological
simulations of $256^3$ particles.  Two typical Cold Dark Matter (CDM)
models have been analyzed, and the halos are selected to have at
least $3\times 10^4$ particles in order to reliably measure the angular
momentum profile. In contrast with the recent claims of Bullock et
al., we find that the degree of misalignment of angular
momentum within a halo is very high. About 50 percent of halos have
more than 10 percent of halo mass in the mass of negative angular momentum
$j$. After the mass of negative $j$ is excluded, the cumulative mass
function $M(<j)$ follows approximately the universal function proposed
by Bullock et al., though we still find a significant fraction of
halos ($\sim 50\%$) which exhibit systematic deviations from the universal
function. Our results, however, are broadly in good agreement with a
recent work of van den Bosch et al..  We also study the angular
momentum profile of halos in a Warm Dark Matter (WDM) model and
a Self-Interacting Dark Matter (SIDM) model. We find that the angular
momentum profile of halos in the WDM is statistically indistinguishable from
that in the CDM model, but the angular momentum of halos in the SIDM
is reduced by the self-interaction of dark matter.
\end{abstract}

\subjectheadings{galaxies:formation-galaxies:structure-galaxies:spiral-cosmology:theory- dark matter}

\section{Introduction}

The formation of galactic disks is one of the most significant unsolved
problems in cosmology. In the popular hierarchical clustering
framework (White \& Reese 1978), disks form in the potential wells of
dark matter halos while the baryons cool and collapse dissipatively
(Fall \& Efstathiou 1980). Assuming that the angular momentum of the gas is
conserved during the collapse and that the gas has the same spin
$\lambda$ parameter as the dark matter in a halo, one can derive a density
profile of cooled gas under an additional assumption that either the
density profile is assumed to be exponential as observed or the angular
momentum of dark matter within a halo follows a distribution,
say, given by a uniform sphere in solid body rotation
(e.g. Fall \& Efstathiou 1980, Blumenthal et al.1986, Dalcanton, Spergel, \& Summers 1997, Jimenez et al.1997, Mo, Mao, \& White 1998; van den Bosch 1998; Avila-Reese \& Firmani 2000). While the theory is quite successful in explaining many observational data of spiral galaxies (e.g. Mo, Mao, \& White 1998), it has encountered
two potentially serious difficulties.

It has been known for many years that detailed numerical simulations
of gas collapse in Cold Dark Matter (CDM) models (Navarro \& Benz
1991; Navarro \& White 1994; Steinmetz \& Navarro 1999;
Navarro \& Steinmetz 2000) have consistently indicated that the
infalling gas loses too much angular momentum due to the dynamical
friction, and the resulting disks are accordingly too small to be
compatible with the observations.  This discrepancy is known as the
``angular momentum catastrophe'' in disk galaxy formation(Navarro, \& Benz
1991). It is unknown if this is a failure of CDM models
(Sommer-Larsen \& Dolgov 2001) or this is due to the simplified
treatments of complicated gas physics in the simulations, e.g. gas
heating from supernova explosion (Weil et al. 1998; Sommer-Larsen,
Gelato \& Vedel 1999; Mo \& Mao 2002).

The second difficulty emerged recently after Bullock et al. (2001;
hereafter B2001) claimed, based on a high-resolution simulation of a
low density flat CDM model (LCDM), that the dark matter halos have a
universal angular momentum profile, i.e. the mass distribution of
specific angular momentum $j$ in a halo obeys a universal form,
\begin{equation}
M(<j)= \frac{M_{\rm v}\mu j}{j_{\rm 0}+
j} \,
\label{eq:jdis}
\end{equation}
where $M_{\rm v}$ is the virial mass, $ j_{\rm 0}=(\mu-1) 
j_{\rm max}$, 
and $j_{\rm max}$
is the maximum specific angular momentum. The parameter $\mu$
indicates the shape of the profile.  Under the conventional
assumption that the angular momentum of the gas is the same as that of
the dark matter before gas collapse and is conserved in detail during
the collapse, the distribution (\ref{eq:jdis}) implies that the formed 
disks contain too much low angular momentum mass to be compatible  with 
the high-resolution observations of galactic disks(Bullock et al. 2001, 
van den Bosch 2001; van den Bosch et al. 2001).

In this work we will examine the validity of equation (\ref{eq:jdis})
using independent simulation data. A large set of cosmological N-body
simulations of the Standard Cold Dark Matter (SCDM) model and a LCDM
model (Jing \& Suto 1998) are employed. The simulations have a similar
resolution to that of B2001, but the sample used in this work is
significantly larger. The large sample enables us to study the angular
momentum profile by selecting a sample of halos which are well
resolved. The two cosmological models will help to answer if the
angular momentum profile of halos depends on cosmological
parameters. We will also analyze the individual halo simulations (Jing
\& Suto 2000) in a Warm Dark Matter (WDM) model (Schaeffer \& Silk
1988, Colombi, Dodelson \& Widrow 1996, Sommer-Larsen \& Dolgov 2001,
Jing 2001, Avila-Reese et al.2001, Bode, Ostriker, \& Turok 2001)
and in a Self-Interaction Dark Matter (SIDM) model (Spergel \&
Steinhardt 2000, Burkert 2000, Dav{\'e} et al. 2001, Yoshida et
al. 2000) to study if the angular momentum profile of halos is
different in different dark matter models. These alternative dark
matter models have been recently proposed to solve possible problems
facing CDM models at sub-galactic scales, i.e. central steep density
profiles of halos and too numerous sub-halos (Moore et al. 1999, Klypin
et al. 1999).

Several similar works had appeared when the present work was near
completion. Knebe et al. (2001) and Bullock et al.(2001b) have
studied the angular momentum profile of halos in
warm dark matter models which are similar to (but not the same as) ours.
van den Bosch et al. (2002) also
examined whether equation (\ref{eq:jdis}) is a good prescription for the
angular momentum profile of halos in an N-body/hydrodynamic
simulation. Since the simulations used by these authors are quite
distinct, in terms of simulation volumes and simulation techniques,
from those used in this paper, our work interestingly complements to
theirs. Wherever necessary, we will compare with their results in the
following sections.

The outline of this paper is as follows. The cosmological models and
the numerical simulations are described in the next section.  The
methods for measuring the angular momentum profile and the results
for the cosmological simulations are presented in section 3. In
section 4, we discuss the results for the WDM and SIDM models. The
conclusions are given in section 5.

\section{Models and numerical simulations}
One set of cosmological simulations used in this paper is from Jing \&
Suto (1998).  Three CDM models were studied by Jing \& Suto (1998),
but here we consider only the SCDM and LCDM models. The SCDM is the
(ever) standard CDM model, and the LCDM is a currently popular flat
low-density model with the density parameter $\Omega_0=0.3$ and the 
cosmological constant $\lambda_0=0.7$. The shape
parameter $\Gamma=\Omega_0 h$ and the amplitude $\sigma_8$ (the rms
top-hat density fluctuation of radius $8\mpc$ at the present time) of
the linear density power spectrum are 0.5 and 0.62 for the SCDM, and 0.2
and 1 for the LCDM. The simulations were generated with a vectorized
$\pppm$ code using $256^3$ simulation particles. The simulation
boxsize is $100\mpc$ (hereafter SCDM100 and LCDM100), and there are
three realizations for each model.  After the work of Jing \& Suto
(1998), another set of simulations has been generated for a boxsize
$50\mpc$ in these CDM models (hereafter SCDM50 and LCDM50). Each model
has two realizations. The resolutions of these simulations are similar
to that used by B2001 (i.e. $256^3$ particles for a simulation box of
$75\mpc$).

The dark matter halo candidates are identified with the
friends-of-friends method (FOF). A linking length $b$ equal to 10\% of
the mean particle separation is adopted. For each candidate halo, we
compute the gravitational potential for each particle and find the
spherical over-density around the particle which has the potential
minimum. The boundary of the candidate halo is determined as the
sphere within which the mean density is equal to the virialization
density $\rho_{\rm v}$. According to the fitting formula of Bryan \&
Norman (1998) for $\rho_{\rm v}$, $\rho_{\rm v}=101 \rho_{\rm crit}$ for the
LCDM model and $\rho_{\rm v}=178 \rho_{\rm crit}$ for the SCDM model, where
$\rho_{\rm crit}$ is the critical density. Some of the candidate halos may
overlap partly, i.e. the separation between two halos is less than the
sum of their virial radii. If two halos are overlapped, the less
massive halo is thrown away. Only the halos with the particle number
$N_{\rm v}$ within the virial radius more than $3\times 10^4$ are included in our
analysis. There are 183,243,307,215 halos in LCDM100, LCDM50, SCDM100 and SCDM50
respectively. For a comparison, B2001 used $\sim 200$ halos with $N_{\rm v}\ge
6\times 10^3$ and $\sim 600$ halos with $N_{\rm v}\ge 10^3$  .

Since the accuracy of the angular momentum measurement critically
depends on the mass resolution (i.e. the number of particles in a
halo, cf. eq.\ref{eq:err}), and the angular momentum of a halo depends
only weakly on its physical mass (as many previous studies of the spin
parameter have shown), we combine the halos in LCDM100 and LCDM50, and
form two samples according to the number of particles in halos. The
first sample, called LCDM-CM, contains 10 halos with more than
$10^5$ particles, and the second one, called LCDM-C, contains all
the halos (426) with more than $3\times 10^4$ particles. Similarly,
two halo samples, called SCDM-CM and SCDM-C, are formed in the SCDM
model, and they contain 27 and 522 halos respectively. A summary of the halo samples are given in Table 2.

In order to study whether the angular momentum of halos depends on
the physical properties of dark matter, we use the sample of Jing
(2001) for our analysis. Jing(2001) used the multi-mass particle code
of Jing \& Suto (2000) to re-simulate 15 halos in a WDM model. The WDM
model has the same model parameters as the LCDM, except that the
linear power spectrum is a WDM type with the free streaming scale 
$R_{\rm f}=0.112\mpc$
(corresponding to a particle mass $700$eV; see Bardeen et al. 1986). The
halos have mass $5\times 10^{12}\himsun$,$6\times 10^{11}\himsun$, and 
$7\times 10^{10}\himsun$ with five halos at each mass.  About $7\times
10^5$ particles are used to simulate the halos with $3\times 10^5$ particles
ended up within the virial radius. For an accurate comparison with 
the LCDM model,
Jing (2001) also simulated 15 halos in the LCDM. The difference
between the 15 LCDM halos and the 15 WDM halos is only the linear
power spectrum, so each pair of the LCDM and WDM halos can be compared
precisely
to show the dependence on the input linear power spectrum. For details about
the simulations, we refer readers to Jing (2001). We call the LCDM
halos as LCDM-H halos, in order to distinguish them with those from
the cosmological simulations.

The Self-Interaction Dark Matter (SIDM) model is the same as the LCDM,
except that the dark matter may be self-interacting. First, five halos
of mass $5\times 10^{12}\himsun$ are selected from our LCDM simulation (Jing \& Suto
1998), and are re-simulated with the multi-mass particle code. The
halos have $\sim 3\times 10^5$ particles
within the virial radius. The self-interactions are implemented in the
code following Burkert (2000) and Dav{\'e} et al. (2000).  The
parameter for the self-interaction is the cross section per unit of
mass $\sigma$. We adopt $\sigma=0$, 0.1, 1.0, and 10 ${\rm cm^2/g}$
for our simulation, and have the 5 halos simulated at each
$\sigma$. 

\section{Results of cosmological simulations}
A first step towards measuring the distribution of specific
angular momentum within a halo is to measure its global
angular momentum. The angular momentum of a halo of N
particles is defined by
\begin{equation}
       {\bf J} =\sum_{i} m_{\rm i} {\bf r}_{\rm i}\times {\bf v}_{\rm i}
\label{eq17}
\end{equation}
where ${\bf r}_{\rm i}$ and ${\bf v}_{\rm i}$ are the position and velocity of
the i-th particle with respect to the halo center of mass. Following
Mo et al. (1998; see also B2001), we measure the spin parameter
$\lambda$ according to the following formula
\begin{equation}
     \lambda = \frac{J}{\sqrt{2} M_{\rm v} 
V_{\rm c} r_{\rm v }}
\label{eq18}
\end{equation}
where $V_{\rm c}$ is the circular velocity at the
virial radius $r_{\rm v}$.

The aim of the paper is to study how the angular momentum is
distributed in a halo. Following B2001, we first determine the global
angular momentum ${\bf J}$ for each halo, and define the z-axis as
pointing to the direction of ${\bf J}$. We then compute the
distribution function of the mass $M(<j)$ which has the specific
angular momentum less than $j$ along the z-axis.  In order to measure
the distribution function, we divide the spherical volume of each halo
into $10$ radial shells, each containing the same number of particles.
Every radial shell is further divided into 6 azimuthal zones of equal
solid angle between $\cos\theta=-1$ and $1$. The two zones with the
same $r$ and $|\cos\theta|$ which are above and below the equatorial
plane are assigned to one ``cell'', and thus every halo is effectively
divided into 30 cells each of which contains approximately the same
number of particles.

The specific angular momentum $j$ of a cell measures the bulk rotation
(motion in the x-y plane) of the particles in the cell. A spin
parameter $\lambda\approx 0.04$ means that the bulk velocity is only a
few percent of the random motion velocity of the particles. Thus, the
discreteness of the particles would be the main source for the
measurement error of $j$ when the number of particles in the cell is
not much more than $10^3$. For a halo of the circular velocity
$v_{\rm c}(r)$ at the radius $r$, the error of $j$ can be estimated by
\begin{equation}
 \sigma_{\rm j}= \frac{r v_{\rm c}(r)}{\sqrt{N}_
{\rm c}}
\label{eq:err}
\end{equation}
where $N_{\rm c}$ is the particle number in the cell, and $r$ is the mean
distance of the cell from the halo center.This is likely an upper limit on the scatter, because the motion of particles is not completely random.

In Figure 1 and Figure 2, we show several examples of the angular
momentum distribution functions in the two cosmological models. Each
halo has more than $10^5$ particles, so the signal-to-noise of the
measurement is high. In the right panels, the cosine of the
angle $\theta$ between the angular momentum vector ${\bf j}$ of a cell
and the z-axis is presented. Among all the $426$ LCDM halos, there are $68$ halos (in the top-right panels) with all cells having $\cos\theta>0$, so the angular momentum of
the different cells aligns very well within these halos. This number is $108$ out of $522$ in the SCDM model. Also in $9$ halos of LCDM model and $12$ halos of SCDM model,(the third
halos from the top), more than 50 \% of the cells have $\cos\theta<0$,
i.e.  the matter within these halos aligns very poorly. But in most
of the halos, (i.e. $253$ out of $426$ halos  in LCDM model and $319$ out of $522$ halos in SCDM model)about 10 to 20 percent of the cells have the negative
angular momentum as the second halos from the top have shown. The
fraction of the mass which has negative $j$ is
significantly higher than B2001 found. In their paper, B2001 claimed 
that ``about 5\% of halos
have a significant amount of their total mass ($\ge 10\%$) contained
in negative $j$ cells''. We note our results are in good agreement with a recent work of 
van den Bosch et al. (2002) who found a similar discrepancy with B2001, though their analysis is for halos of galactic mass and at redshift $z=3$. The
percentage of the matter having negative $j$ will be discussed in more
detail below. Since the fraction of negative $j$ cells is
non-negligible, the universal function (\ref{eq:jdis}) certainly
fails for describing most of our halos, for the fitting formula
assumes that the mass of anti-aligned $j$ is negligible. Following
B2001, we exclude all cells of negative $j$ and compare the $j$
distribution function with equation (\ref{eq:jdis}). In this case, we
replace the virial mass $M_{\rm v}$ in the equation with the total mass of
positive $j$ cells $M(j>0)$. A comparison of our simulation result with
the modified formula is presented in the left panels of the
figures. Considering that the measurement of $j$ in every cell has an
error, when the cumulative mass $M(<j)$ is considered, the error for
the upper limit $j$ might be underestimated if Eq.(\ref{eq:err}) is used. Conservatively, the error
bars for the upper limit $j$ in the figures are calculated by
$\sigma_{<{\rm j}}^2=\sum_{\rm j1\le \rm j} \sigma_{\rm j1}^2$. From the figure, it can be
easily seen that the simulation data can be reasonably ($\sim 50 \%$) described by
equation(\ref{eq:jdis}), but not perfectly. As the figures show, a
significant fraction of halos has a higher fraction of low $j$ mass 
than the fitting formula. Because the observational data of disk
galaxies seem to indicate less low $j$ mass than the function
(\ref{eq:jdis}) (B2001, van den Bosch et al. 2001, van den Bosch
2001), our results imply that the discrepancy between the CDM models
and the observations could be more serious unless some (unclear)
physical mechanisms are in play to modify the $j$ distribution of
the gas which forms the discs (e.g. van den Bosch et al. 2002; Maller
\& Dekel 2002). Only a very small fraction ($15$ halos for LCDM model and $18$ halos for SCDM model) of halos (e.g. the
bottom examples in the figures) have less low $j$ mass than the
formula (\ref{eq:jdis}). These halos may be in a better agreement with
the observational data of disk galaxies, but considering that the disk
galaxies are common, it is unlikely that such halos are numerous enough 
 to explain the observations.

In order to quantify how much mass in halos is contained in negative
$j$ cells, we present in Figure 3 the percentage of the halos which
have a fraction of mass $M(j<0)/M_{\rm v}$ in negative j cells. The
upper panel shows for LCDM-H halos which have more than $3\times
10^{5}$ particles. Among the $15$ halos, $12$ halos have negative j
cells and $4$ halos have $M(j<0)/M_{\rm v}$ higher than
10\%.In the middle panel, we show the results for LCDM-C (the solid
line) and LCDM-CM (the dotted line) samples. From the figure, it can
be easily read that about 40\% of LCDM-C halos and 55\% of LCDM-CM
halos have $M(j<0)/M_{\rm v}> 0.1$. Similarly, the percentage of the
halos with $M(j<0)/M_{\rm v}> 0.1$ is 50\% in the SCDM-C and 60\% in the
SCDM-CM sample. We have
further used the Kolmogorov-Smirnov test to make sure if there is any
significant difference in the misalignment distribution between the
different (sub)-samples. The test results, listed in Table 1, show
that the distribution functions of the three subsamples in the LCDM
model are consistent with being drawn from the same parent
distribution. The distributions of the two subsamples in the SCDM
model are also consistent. Therefore, the misalignment could not be a result from the discreteness effect in the simulations.
Only significant difference has been found
between SCDM-C and LCDM-C, since the both samples are large and there
is significantly more misalignments of $j$ in the SCDM-C.

The misalignments of $j$ could be caused by the substructures in the
halos. Since there is a higher amount of substructures in the
SCDM than in the LCDM model (e.g. Jing et al. 1995), we can easily
explain why the SCDM-C halo contain a higher fraction of negative
$j$ mass than the LCDM-C halos. Following Jing (2000; hereafter
J2000), we use the clumpness of the density profile as an indicator
for the substructures in halos. The clumpness is quantified by the
maximum relative deviation ${\rm dvi}_{\rm max}$ of the simulation density
profile $\rho(r_{\rm i})$ from the fitting $\rho_{\rm NFW}
(r_{\rm i})$ of the
Navarro-Frenk-White (1996) form:
\begin{equation}
 {\rm dvi}_{\rm max}={\rm max}\{\vert \frac{{\rm \rho}(r_
{\rm i})-\rho_{\rm NFW}(r_{\rm i})}
{\rho_{\rm NFW}(r_{\rm i})}\vert\}
\label{eq21}
\end{equation}
where $i$ runs over all radial bins of the density profile.  Figure 4
plots the fraction of negative $j$ mass in halos as a function of the
substructure indicator ${\rm dvi}_{\rm max}$. Since most of the halos ($401$ out of $426$ LCDM halos and $501$ out of $522$ SCDM halos) have 
${\rm dvi}_{\rm max}\leq0.6$, from the figure we find that the misalignment
of $j$ is nearly independent of the substructures. The peaks at ${{\rm dvi}_{\rm max}=1.05}$ in the LCDM model and at ${{\rm dvi}_{\rm max}=0.75}$ in the SCDM model may be contributed by statistical fluctuations, for there is only one halo in the LCDM peak and $7$ halos in the SCDM peak.

We have also examined where the misalignment of $j$ happens in
halos. Figure 5 shows the percentage of negative $j$ cells at
different radius $r$. The results indicate that either at small or at
large radii, there are more mass with negative $j$ than at the median
radius.The result can be easily understood, since the global angular
momentum of a halo is mainly determined by the mass at the median
radius, and the mass at the median radius will naturally align better 
with {\bf J}. This however indicates that the twisting of the angular
momentum vectors occur everywhere inside halos.

It may also be interesting to study the distribution of the shape
parameter $\mu$ in the function $M(<j)$ (Eq.\ref{eq:jdis}), and
compare our results with B2001.  We should emphasize again that only
positive $j$ cells are included in the fitting to our
simulation results. In the function, we have replaced $M_v$ in B2001
with $M(j>0)$.  The $\mu-1$ parameter follows a log-normal
distribution (Figures 6 and 7) as what Bullock et al.(2001) found. The mean values
$\bar \mu$ and the scatter $\sigma_{\ln(\mu-1)}$ are given in Figure 8
for LCDM and SCDM. From these two parameters, we can
conclude that the distribution of $\mu$ does not depend on the spin
parameter $\lambda$. In the SCDM model $\bar \mu =1.27$ and
$\sigma_{\ln(\mu-1)}=0.41$, and in the LCDM model $\bar \mu =1.22$ and
$\sigma_{\ln(\mu-1)}=0.41$. We note that there is a tendency of the
increase of $\mu-1$ with the spin parameter $\lambda$ in B2001, which
has not been confirmed by our analysis.

\section{Alternative dark matter models}
As already pointed out in the previous section, the CDM models have
the difficulty to explain the observational data of disk galaxies
under the conventional assumption that the angular momentum
distribution of gas follows that of dark matter. While the assumption
of gas tracing dark matter
may well be wrong when realistic but complicated physical processes of
galaxy formation are taken into account, we study here the angular
momentum distribution in the Warm Dark Matter (WDM) model and in the
Self-Interacting Dark Matter (SIDM) model. We want to see if a change 
of dark matter species can rescue the hierarchical galaxy formation models.

In Figure 9, we present a comparison of the angular momentum
distribution between LCDM-H halos and WDM halos. The mass of 
the halos are $6\times 10^{11}\himsun$
and $7\times 10^{10}\himsun$. From Jing (2001),
the streaming effect of WDM is significant at these mass scales.In the top two
panels, $M(<j)/M(j>0)$ of two LCDM-H halos are presented (triangles and
the solid line) together with those of the corresponding WDM halos
(open circles and the dotted line). The difference in $M(<j)/M(j>0)$
between the two dark matter models is significant, but is not
systematic. In the low panels, we show comparisons of the fitting
parameter $\mu$ and the spin parameter $\lambda$ between LCDM-H and
WDM halos. The results clearly show that there is a correlation of the
parameters between the two models but the scatter is quite
large. The WDM halo may contain a lower (top left) or higher
(top right) amount of low $j$ mass than the corresponding LCDM halo.
There is however no indication for any systematic difference in
the angular momentum distribution between the two dark matter
models. Our results agree well with the recent studies of Knebe et
al. (2001) and Bullock et al. (2001b).

The distribution of the angular momentum of halos in SIDM models has
similar profiles to those of the LCDM model. Figure 10 presents a
comparison of the fitting parameter $\mu$ and the spin parameter
$\lambda$ between the LCDM ($\sigma=0$) and SIDM halos. There is a
tendency that both $\mu$ and $\lambda$ become smaller in the models of
stronger interaction (i.e. larger $\sigma$), i.e. the angular momentum
is reduced by the dark matter interaction. This indicates that SIDM models may
have more difficulties to reproduce the observation data of disk
galaxies than the conventional LCDM models.

\section{Discussion and conclusions}
We have studied the angular momentum profile of dark matter halos for
a statistical sample drawn from a set of high-resolution cosmological
simulations.  Two typical CDM models have been considered, and the
halos are selected to have at least $3\times10^4$ particles in order to
reliably measure the angular momentum profile. In contrast with the
recent claims of B2001, we find that the degree
of misalignment of the angular momentum within a halo is very
high. About 50 percent of halos have more than 10 percent of halo mass
in the mass of negative angular momentum $j$. After the mass of negative $j$
is excluded, the cumulative mass function $M(<j)$ follows
approximately the universal function proposed by B2001 if the virial
mass $M_{\rm v}$ in the function is replaced by $M(j>0)$, though we still
find a significant fraction of halos ($\sim 50\%$) which exhibit
systematic deviations from the universal function. Our results,
however, are broadly in good agreement with the recent work of van den
Bosch et al. (2002).

We have also studied the angular momentum profile of halos in the Warm
Dark Matter model and in the Self-Interacting Dark Matter model in order to
study how the angular momentum profile is affected by the basic
assumption about the dark matter. We have made a detailed comparison
between the halos in these scenarios and the corresponding halos in the
LCDM model. We find that there is no {\it systematic} difference in
the angular momentum between the halos from the WDM and from the LCDM,
though a pair of corresponding WDM and LCDM halos may exhibit quite
different angular momentum profiles. We also find that
the self-interaction of dark matter in the SIDM models can generally
reduce the angular momentum, which makes the spin parameter $\lambda$
and the shape parameter $\mu$ smaller. Thus it seems that these
dark models do little help to solve the angular momentum problem
encountered by the CDM models.

Our results also indicate that it should be cautious to use the
universal angular momentum profile of B2001 to predict observational
properties for disk galaxies. The angular momentum in different parts
of a halo does not orient as coherently as B2001 claimed. The mass of
negative angular momentum $j$ may combine with those mass of small $j$
to form the bulge component in spiral galaxies (van den Bosch et
al. 2002), or the angular momentum profile of the gas in a halo is
significantly different from that of the dark matter due to
hydro-dynamical processes like heating and explosions (Maller \& Dekel
2002). The relation between the disk properties and the angular
momentum profile (Eq.\ref{eq:jdis}) could be more complicated than
previously thought.

\acknowledgments 
We would like to thank van den Bosch for useful discussion. The work is
supported in part by the One-Hundred-Talent Program, by
NKBRSF(G19990754) and by NSFC(No.10125314).


\newpage
\begin{figure}
\epsscale{1.0} \plotone{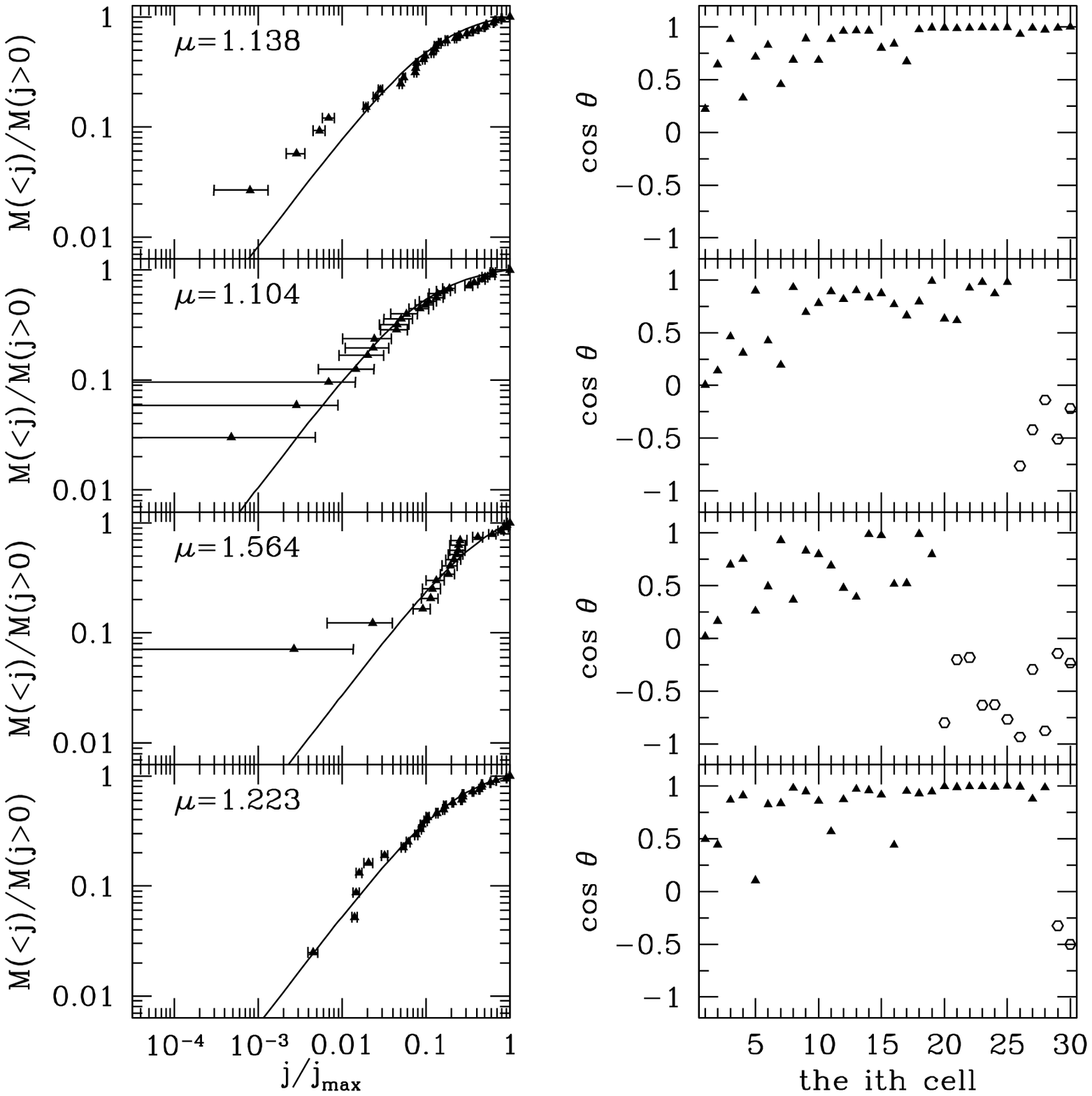}
\caption{{\it Left panel} -- the mass distribution of specific angular
momentum of four halos in the LCDM model. {\it Right panel} -- the
$\cos\theta$ of the cells in the corresponding halos. Solid and open
symbols are used for positive and negative values of $\cos\theta$
respectively.}
\end{figure}
\begin{figure}
\epsscale{1.0} \plotone{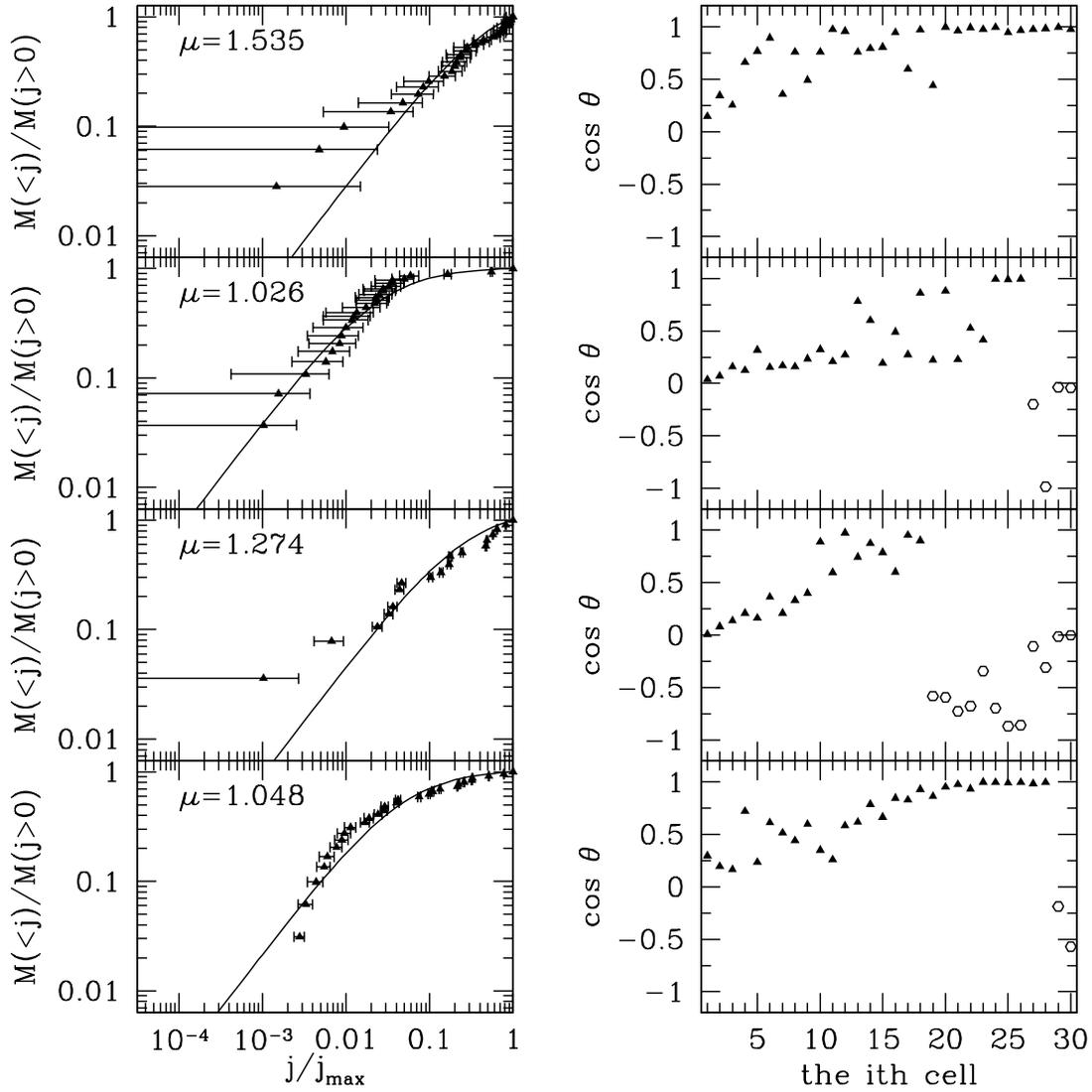}
\caption{The same as Fig.~1, but for four halos in the SCDM model.}
\end{figure}
\begin{figure}
\epsscale{1.0} \plotone{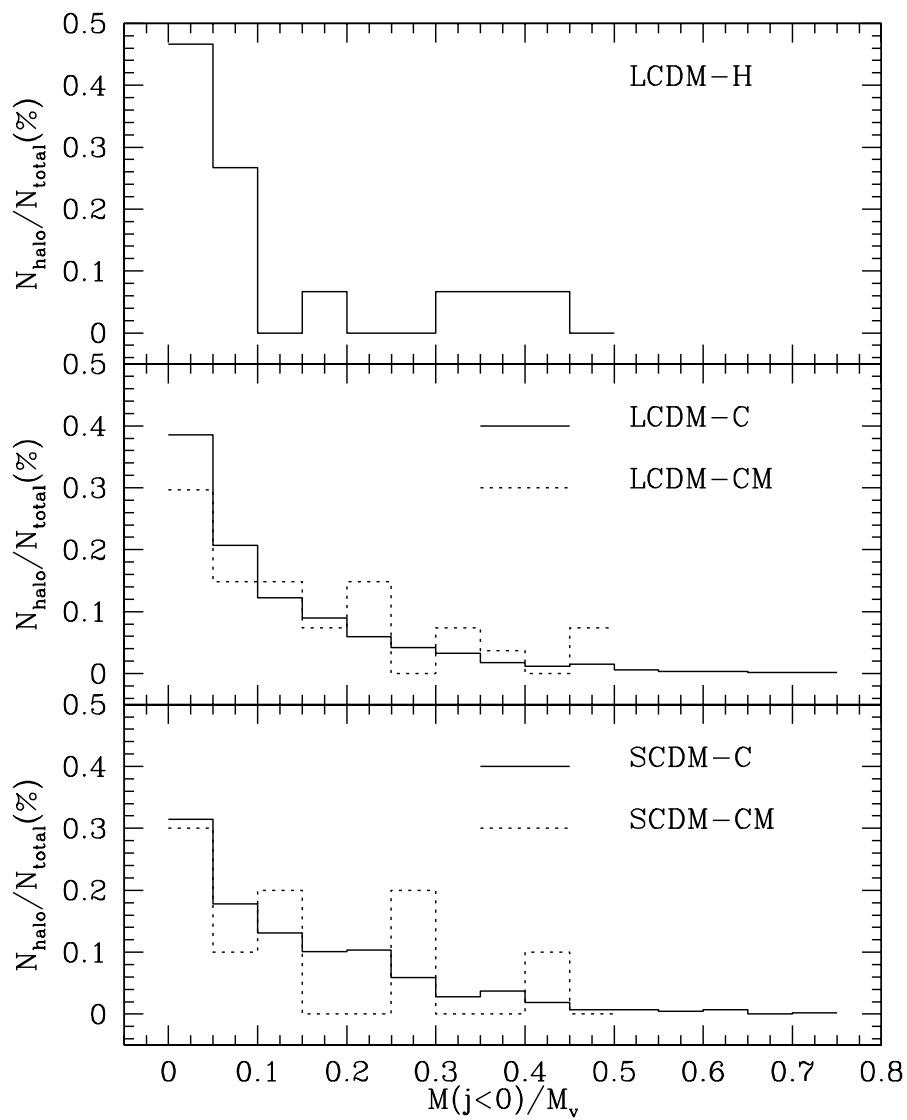}
\caption{The percentage of halos having mass $M(j<0)$ contained in
negative $j$ cells. LCDM-H halos are shown in the top panel. In the
middle panel, the solid histogram is for the LCDM-C halos and the dot
one for the LCDM-CM halos. In the bottom panel, the solid and dotted
histograms are for SCDM-C and SCDM-CM respectively}
\end{figure}

\begin{figure}
\epsscale{1.0} \plotone{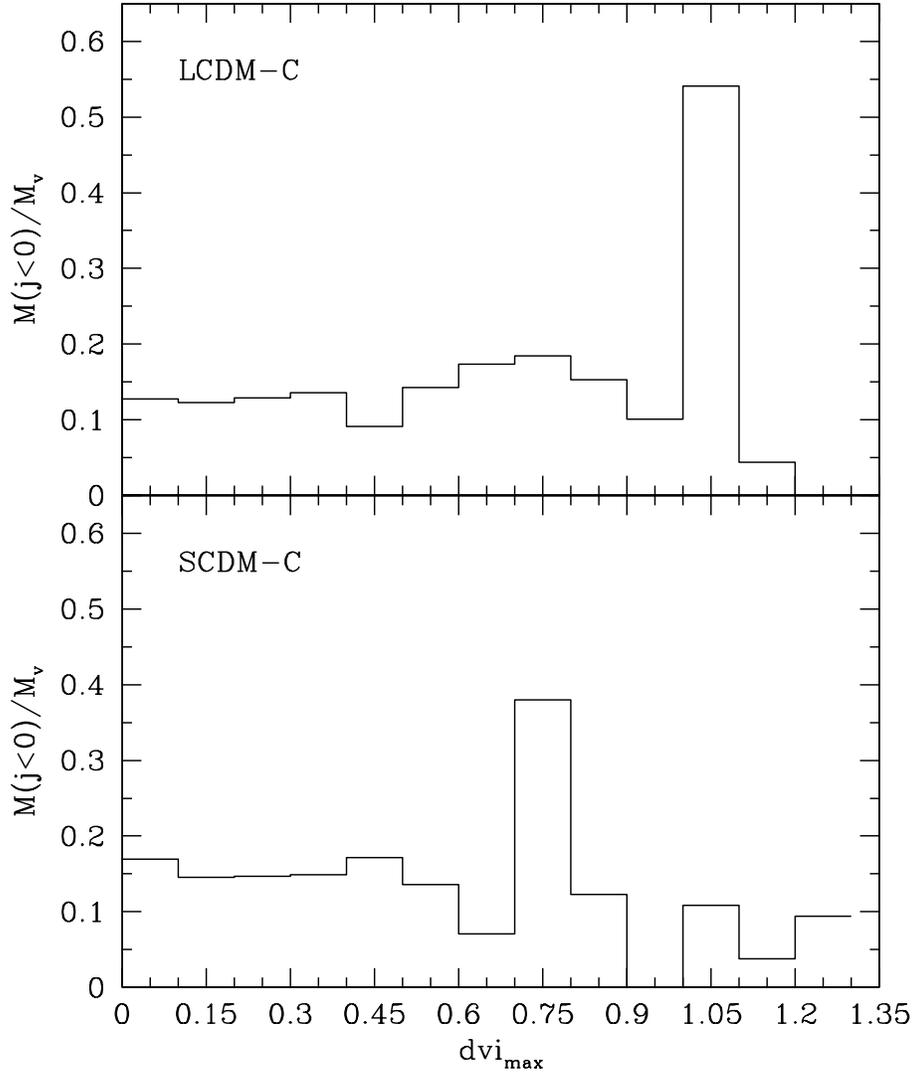}
\caption{The total mass of negative $j$ cells vs the substructure
indicator ${\rm dvi}_{\rm max}$ for both LCDM and SCDM models.}
\end{figure}

\begin{figure}
\epsscale{1.0} \plotone{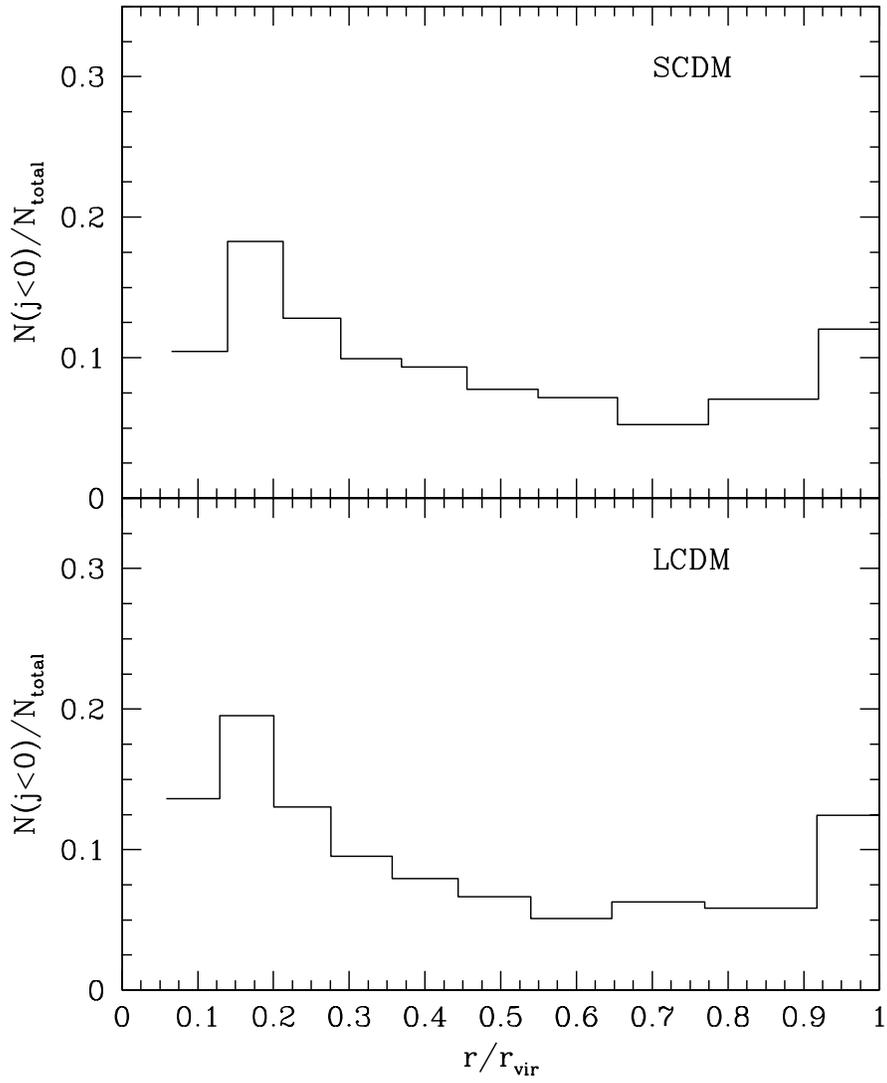}
\caption{The percentage of negative $j$ cells at radius $r$.}
\end{figure}

\begin{figure}
\epsscale{1.0} \plotone{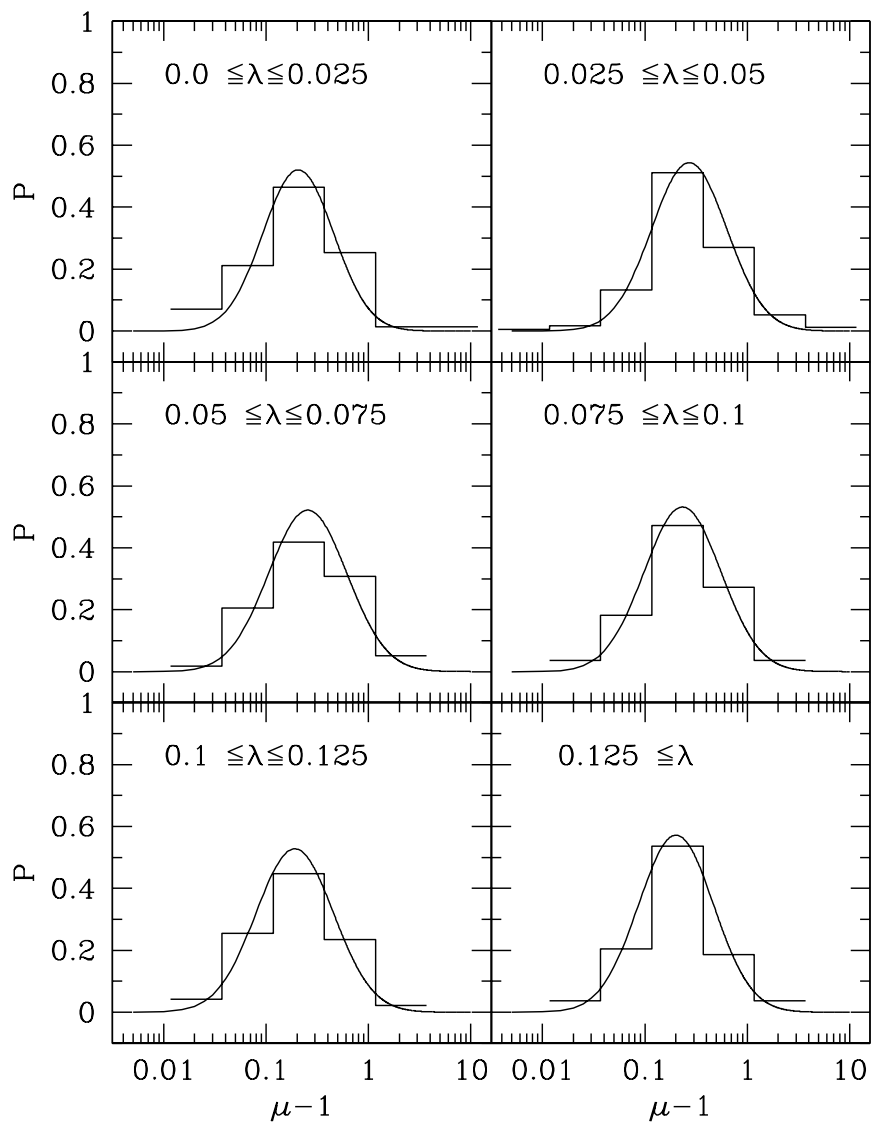}
\caption{The $\mu$ distribution of halos for LCDM model(histogram).
Halos were divided into six groups according to their $\lambda$ values.
The solid curves in each panel are the log-normal distributions of $\mu-1$
with the means and the standard deviations shown in Fig.8} 
\end{figure}
\begin{figure}
\epsscale{1.0} \plotone{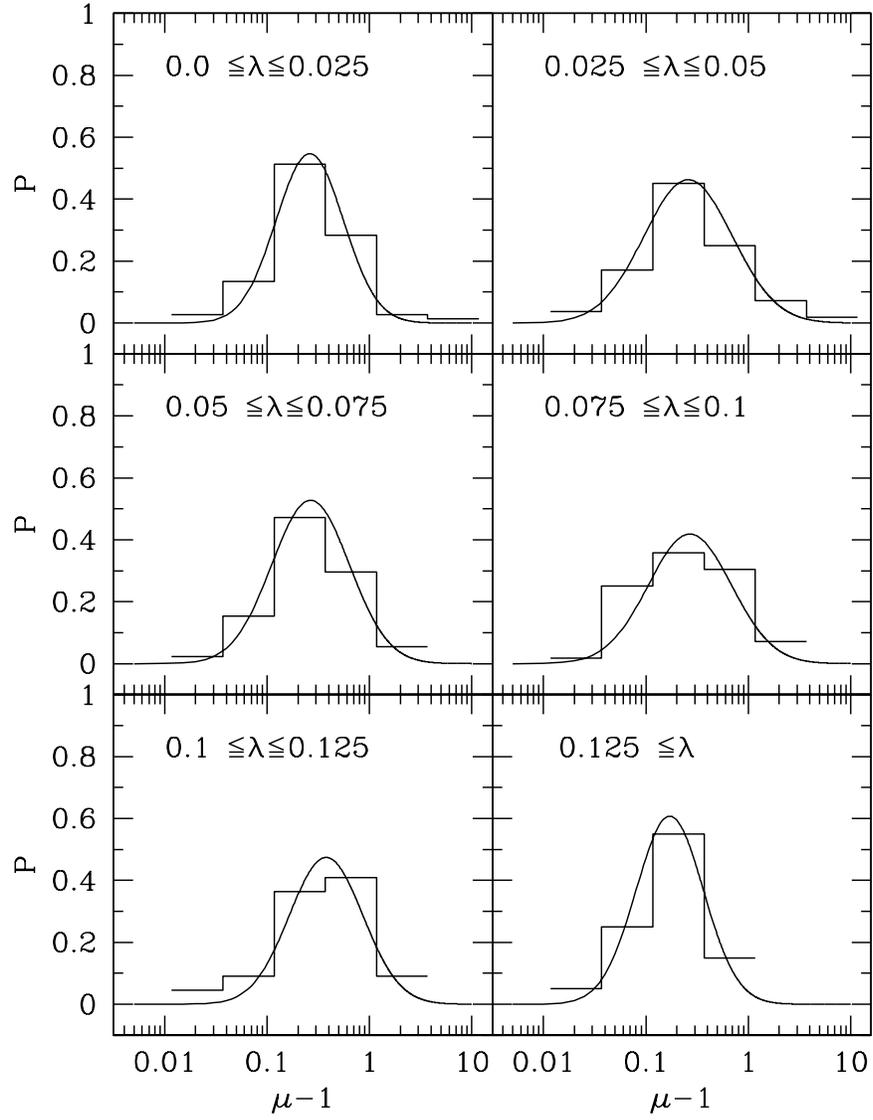}
\caption{The same as Fig.6 but for the SCDM model.}  
\end{figure}
\begin{figure}
\epsscale{1.0} \plotone{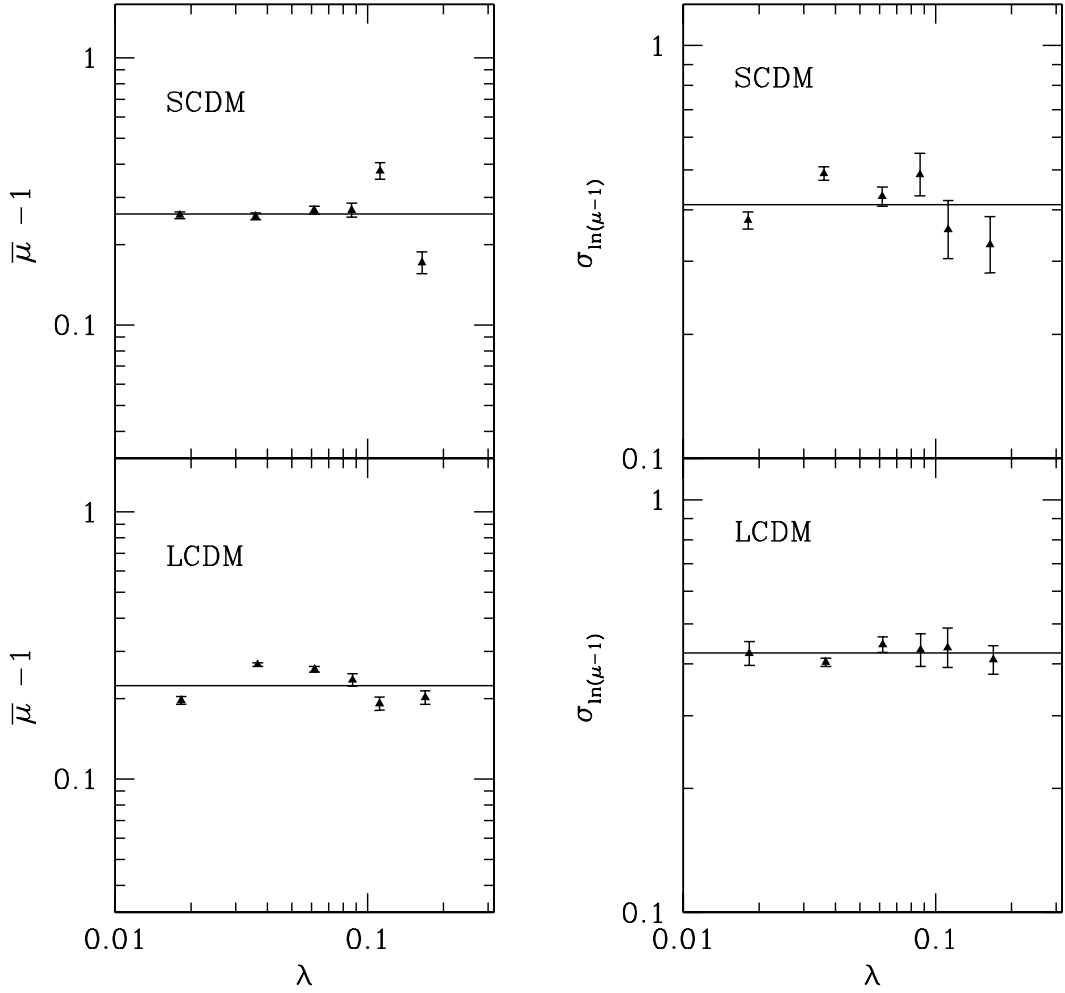}
\caption{The means and standard deviations of the log-normal distribution 
of $\mu -1$ for both LCDM and SCDM model as a function of  $\lambda$.
The solid lines are the average values.}
\end{figure}
\begin{figure}
\epsscale{1.0} \plotone{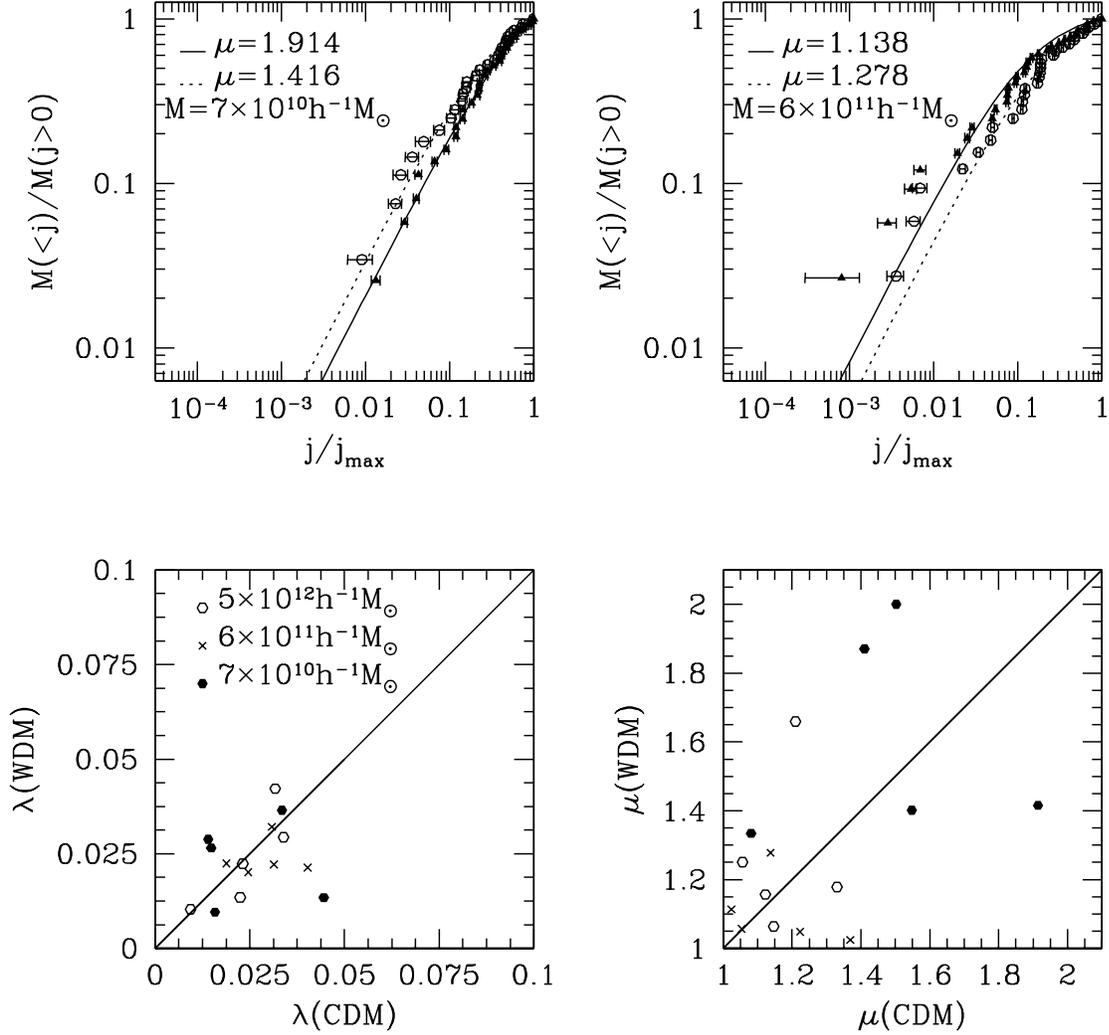}
\caption{ {\it Upper panels} -- The cumulative mass $M(<j)$ as a
function of the specific angular momentum $j$ for two pairs of LCDM
and WDM halos. Open circles and dotted lines are for the WDM halos, and
filled triangles and solid lines for LCDM halos. {\it Lower
panels}--The spin parameter $\lambda$ and shape parameter$\mu$ of the
WDM halos vs. those of the LCDM halos.}
\end{figure}

\begin{figure}
\epsscale{1.0} \plotone{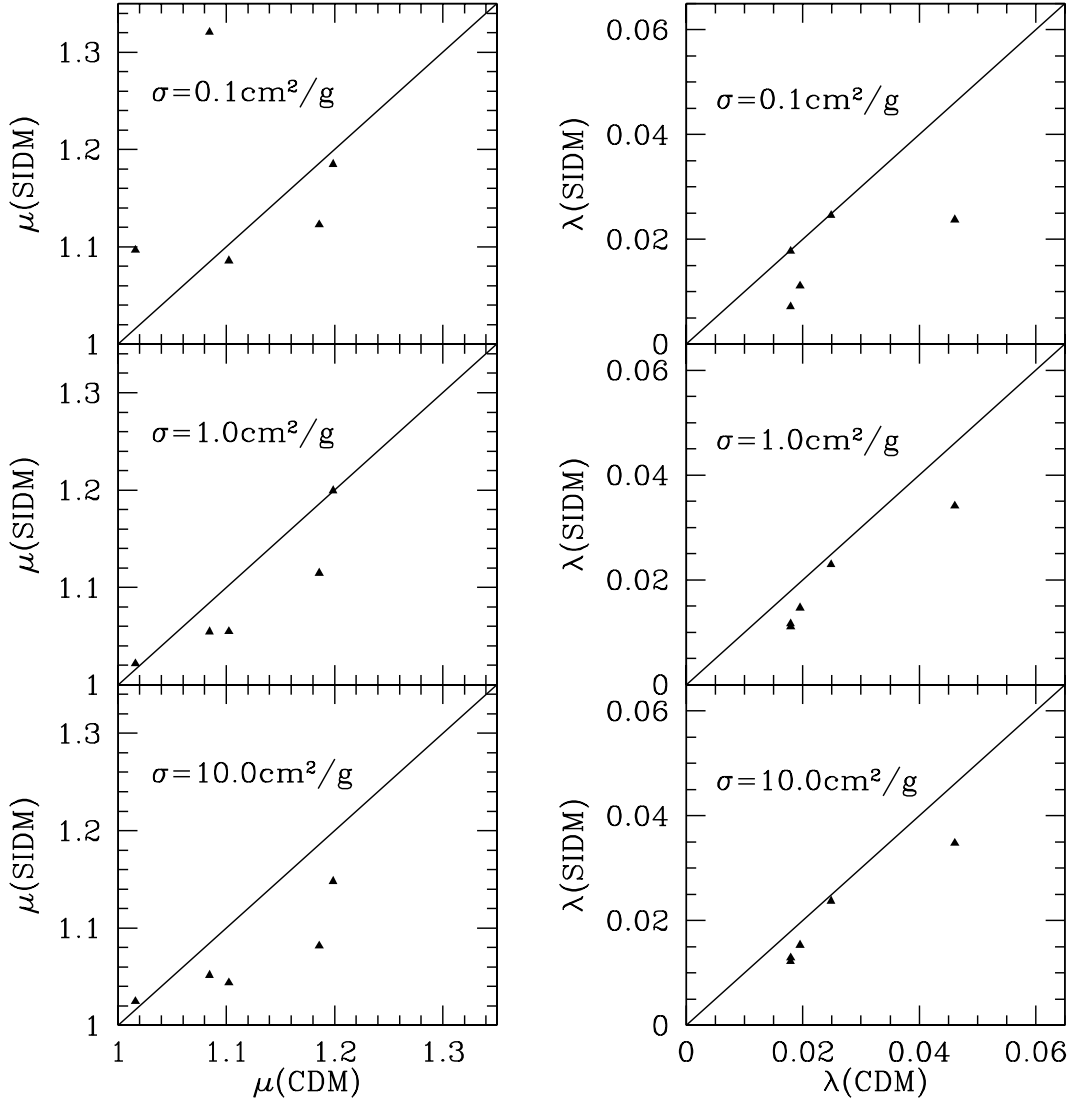}
\caption{The spin parameter $\lambda$ and shape parameter$\mu$ of the
SIDM halos vs. those of the LCDM halos.}
\end{figure}

\newpage
\begin{table}
\caption{The probability of two j-distributions drawn from the same parent distribution}
\begin{center}
\begin{tabular}{cc}
\hline\hline
CDM Model &  Probability \\
\hline
LCDM-H vs LCDM-CM & 0.3374 \\
LCDM-H vs LCDM-C  & 0.8233 \\
LCDM-H vs SCDM-CM & 0.1535 \\
LCDM-H vs SCDM-C  & 0.2261 \\
LCDM-CM vs SCDM-CM  & 0.8441 \\
LCDM-C vs SCDM-C & 0.0026 \\
\hline\hline
\end{tabular}
\end{center}
\end{table}

\begin{table}
\caption{Summary of the simulation models}
\begin{center}
\begin{tabular} {ccccc}
\hline\hline
Model & $L_{box}(\mpc)$ & $m_{p}(\himsun)$ & $N_{v}$\tablenotemark{a} & 
Number of halos \\
\hline

LCDM-H\tablenotemark{b}&  & & $3\times 10^{5}$ & $15$ \\
LCDM-C & 100(LCDM100) & $5.0\times 10^{9}$ & $3\times 10^{4}$ & $426$ \\
 & 50(LCDM50)& & & \\       
LCDM-CM & 100(LCDM100) & $5.0\times 10^{9}$ & $10^{5}$ & $10$ \\
 & 50(LCDM50)& & & \\
SCDM-C & 100(SCDM100) & $1.7\times 10^{10}$ & $3\times 10^{4}$ & $522$ \\
 & 50(SCDM50)& & & \\
SCDM-CM & 100(SCDM100) & $1.7\times 10^{10}$ & $10^{5}$ & $27$ \\
 & 50(SCDM50)& & & \\
\hline\hline
\tablenotetext{a}{The minimum number of particles per halo}
\tablenotetext{b}{The boxsize and mass of the particles are varied} 
\end{tabular}
\end{center}
\end{table}
\end{document}